\documentclass[english,aps,prd]{revtex4}
\usepackage[T1]{fontenc}
\usepackage[latin1]{inputenc}
\usepackage{graphicx}

\makeatletter

\providecommand{\tabularnewline}{\\}

\usepackage{babel}
\makeatother
\begin{document}

\title{Single production of excited electrons at future $e^+e^-$, $ep$
and $pp$ colliders}

\author{O. \c{C}ak{\i}r}

\email{ocakir@science.ankara.edu.tr}

\author{A. Y{\i}lmaz}

\affiliation{Ankara University, Faculty of Sciences, Department of Physics, 06100,
Tandogan, Ankara, Turkey }

\author{S. Sultansoy}

\email{saleh@gazi.edu.tr}

\affiliation{Gazi University, Faculty of Arts and Sciences,
Department of Physics, 06500, Teknikokullar, Ankara, Turkey}

\affiliation{Azerbaijan Academy of Sciences, Institute of Physics, H. Cavid Ave.
33, Baku, Azerbaijan.}

\begin{abstract}
We analyzed the potential of the LC with $\sqrt{s}=0.5$ TeV, LC$\otimes$LHC
based ep collider with $\sqrt{s}=3.74$ TeV and the LHC with $\sqrt{s}=14$
TeV to search for excited electrons through transition magnetic type
couplings with gauge bosons. The $e^{\star}\rightarrow e\gamma$ signal
and corresponding backgrounds are studied in detail.
\end{abstract}

\pacs{12.60.Rc, 13.10.+q, 13.35.-r}

\maketitle

\section{introduction}

Three types of colliders related to the energy frontiers in
particle physics research seem to be promising in the next decade.
Namely, they are Large Hadron Collider (LHC) with the center of
mass energy $\sqrt{s}=14$ TeV and luminosity
$L=10^{34}-10^{35}$cm$^{-2}$s$^{-1}$, linear $e^+e^-$ collider
(LC) with $\sqrt{s}=0.5$ TeV and
$L=10^{34}-10^{35}$cm$^{-2}$s$^{-1}$, and linac-ring type ep
collider (LC$\otimes$LHC) with $\sqrt{s}=3.74$ TeV and
$L=10^{31}-10^{32}$cm$^{-2}$s$^{-1}$ (see \cite{1} and references
therein). Even though the last one has a lower luminosity it can
provide better conditions for investigations of a lot of phenomena
comparing to LC due to the essentially higher center of mass
energy and LHC due to more clear environment. For this reason,
different phenomena (compositeness, SUSY, etc.) should be analyzed
taking into account all three types of colliders.

The fundamental questions left open by the Standard Model (SM),
such as the number of fermion families, the fermion masses and the
mixings, are addressed by composite models \cite{2}. In the
framework of composite models of quarks and leptons, constituents
of known fermions interact by means of new interactions. A
non-trivial substructure of known fermions leads to a rich
spectrum of excited states \cite{3}. Phenomenologically, an
excited lepton is defined to be a heavy lepton which shares
leptonic quantum numbers with one of the existing leptons. Charged
($e^{\star},\mu^{\star}\textrm{ and }\tau^{\star}$) and neutral
($\nu_{e}^{\star},\nu_{\mu}^{\star}\textrm{ and
}\nu_{\tau}^{\star}$) excited leptons are predicted by composite
models where leptons and quarks have substructure.

Current limits on the mass of the excited electron are \cite{4}:
$m_{\star}>100$ GeV from LEP (pair production) assuming $f=f'$
\cite{5}, $m_{\star}>223$ GeV from HERA (single production)
assuming $f=f'=\Lambda/m_{\star}$ \cite{6} and $m_{\star}>310$ GeV
from LEP (indirect) assuming $\lambda_{\gamma}=1$ \cite{7}.

On the theoretical side, the production of excited leptons was
studied at LEP and HERA energies \cite{8} and at hadron colliders
\cite{9} by taking into account only the signal. The LEP and HERA
bounds on excited lepton masses are low enough when compared to
the expected scale for the compositeness. This motivates us to
reanalyze the excited lepton production at the future colliders.
In this paper, we go beyond by studying the signal as well as the
background (with the interference between them) at the similar
experimental conditions and compare the potential of each type of
colliders to search for the single production of excited
electrons.

The interaction between an excited lepton, gauge bosons and the SM
leptons is described by $SU(2)\times U(1)$ invariant lagrangian
\cite{8,9,10}

\begin{equation}
L=\frac{1}{2\Lambda}\bar{l_{R}}^{\star}\sigma^{\mu\nu}\left[fg\frac{\overrightarrow{\tau}}{2}\cdot\overrightarrow{W}_{\mu\nu}+f^{'}g^{'}\frac{Y}{2}B_{\mu\nu}\right]l_{L}+h.c.\end{equation}
where $\Lambda$ is the scale of the new physics responsible for the
existence of excited leptons; $W_{\mu\nu}$ and $B_{\mu\nu}$ are
the field strength tensors; $\overrightarrow{\tau}$ denotes the Pauli
matrices, $Y=-1/2$ is the hypercharge; $g$ and $g^{'}$ are the
SM gauge couplings of SU(2) and U(1), respectively; the constants
$f$ and $f^{'}$ are the scaling factors for the corresponding gauge
couplings. In these expressions, $\sigma^{\mu\nu}=i(\gamma^{\mu}\gamma^{\nu}-\gamma^{\nu}\gamma^{\mu})/2$
where $\gamma^{\mu}$ are the Dirac matrices.

\begin{figure}
\includegraphics[scale=0.8]{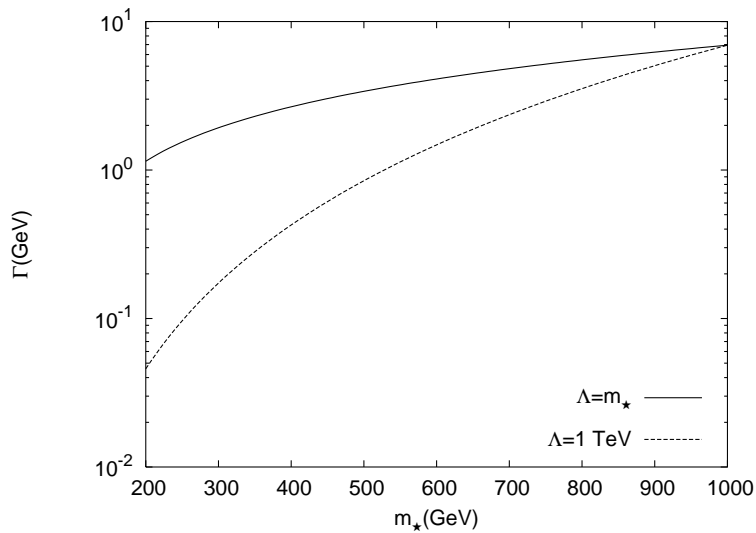}

\begin{center}(a)\end{center}

\includegraphics[scale=0.8]{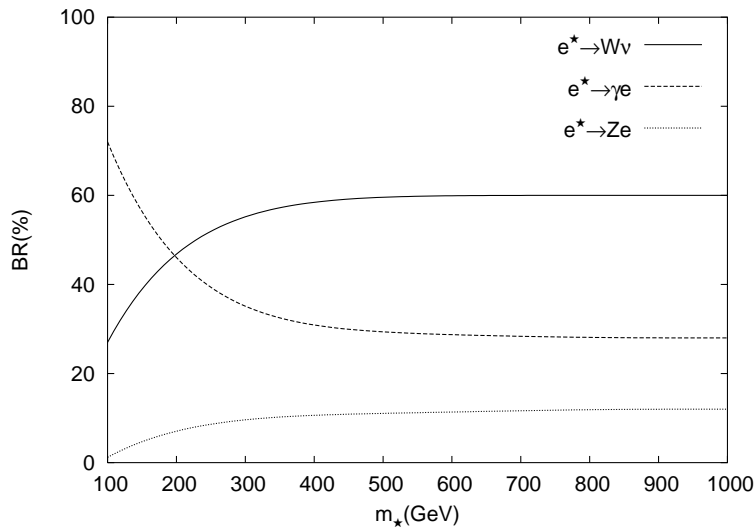}

\begin{center}(b)\end{center}

\caption{(a) The total decay width $\Gamma$ in GeV of excited electron for
the scale $\Lambda=1$ TeV and $\Lambda=m_{\star}$ with the couplings
$f=f'=1$, and (b) the branching ratios BR (\%) depending on the mass
of excited electron for $f=f'=1$.\label{fig1}}
\end{figure}

For an excited electron, three decay modes are possible: radiative
decays $e^{\star}\rightarrow e\gamma$, charged current decays
$e^{\star}\rightarrow\nu W$, neutral currents decays
$e^{\star}\rightarrow eZ$. Neglecting ordinary lepton masses the
decay widths are obtained as \cite{9,10}
\begin{equation}
\Gamma(e^{\star}\rightarrow lV)=\frac{\alpha m_{\star}^{3}}{4\Lambda^{2}}f_{V}^{2}\left(1-\frac{m_{V}^{2}}{m_{\star}^{2}}\right)^{2}\left(1+\frac{m_{V}^{2}}{2m_{\star}^{2}}\right)\end{equation}
where $f_{\gamma}=-(f+f^{^{'}})/2$, $f_{W}=f/(\sqrt{2}sin\theta_{W})$
and $f_{Z}=(-fcos^{2}\theta_{W}+f^{'}sin^{2}\theta_{W})/2$. The total
decay width $\Gamma$ of the excited electron and the relative branching
ratios (BR) into ordinary leptons and gauge bosons $\gamma,Z,W$ are
given in Fig. \ref{fig1}. For a comparison, in Fig. \ref{fig1}(a)
we show the total decay widths of excited electron for $\Lambda=m^{\star}$and
$\Lambda=1$ TeV, which are commonly used for the new physics scale.
For large values of the excited electron mass, the branching ratio
for the individual decay channels reaches to the constant values 60\%
for the $W$-channel, 12\% for the $Z$-channel and 28\% for the photon
channel. The branching ratios in these different modes depend on the
relative values of $f$ and $f^{'}$. For $f=f^{'}$, the radiative
decay is allowed for excited electron whereas it is forbidden for
excited neutrino.

\section{Single production of excited electrons}

We analyze the potentials of the LC, LC$\otimes$LHC and LHC machines
to search for excited electrons (or positrons) via the single production
reactions

\begin{equation}
e^{+}e^{-}\rightarrow e^{\pm\star}e^{\mp}\end{equation}

\begin{equation}
e^{-}p\rightarrow e^{-\star}q(\bar{q})X\end{equation}

\begin{equation}
pp\rightarrow e^{\pm\star}e^{\mp}X\qquad\textrm{and\qquad}pp\rightarrow e^{-(+)\star}(\overline{\nu})\nu X\end{equation}
with subsequent decay of excited electron (or positron) into photon
and an electron (or positron). Therefore, we deal with the process
$e^{+}e^{-}\rightarrow\gamma e^{\pm}e^{\mp}$, and subprocesses $e^{-}q(\bar{q})\rightarrow\gamma e^{-}q(\bar{q})$,
$q\bar{q}\rightarrow\gamma e^{\pm}e^{\mp}$ and $q\bar{q}'\rightarrow\gamma e^{-(+)}(\bar{\nu})\nu$.
The signal and background were simulated at the parton level by using
the program CompHEP 4.2 \cite{11} (the interference terms between
signal and background processes are included). In our calculations
we used the parton distribution functions library CTEQ6L \cite{12}
with the factorization scale $Q^{2}=\hat{s}$.

\begin{figure}
\includegraphics{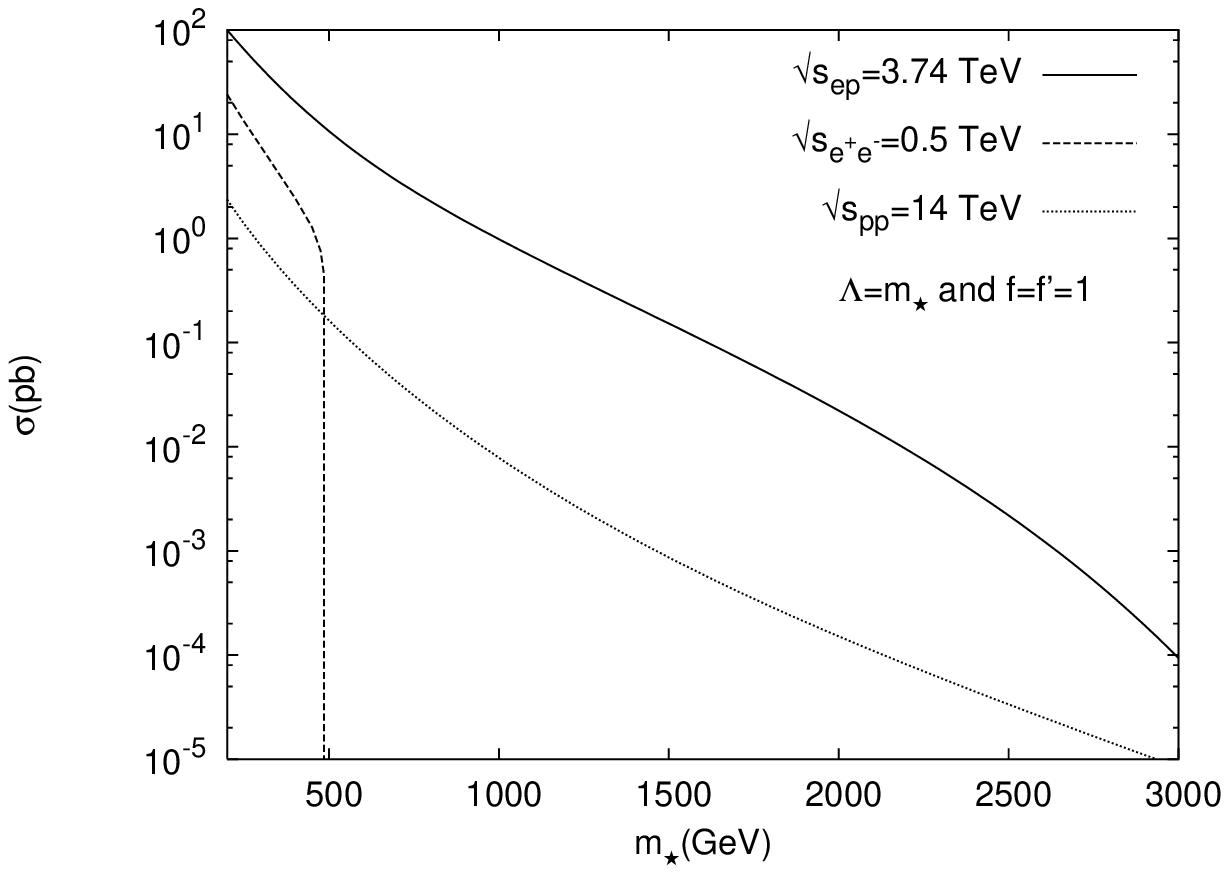}

\caption{The total cross sections for the single production of excited electron
at $e^{-}e^{+}$ collider with $\sqrt{s}=0.5$ TeV, $ep$ collider
with $\sqrt{s}=3.74$ TeV and $pp$ collider with $\sqrt{s}=14$ TeV.
\label{fig2}}
\end{figure}

For a comparison of different colliders, the signal cross sections
for the processes given above are presented in Fig. \ref{fig2}
assuming the scale $\Lambda=m_{*}$ and the coupling parameters
$f=f^{'}=1$.

\subsection{$e^{-}e^{+}$ Collider}

\begin{figure}
\includegraphics{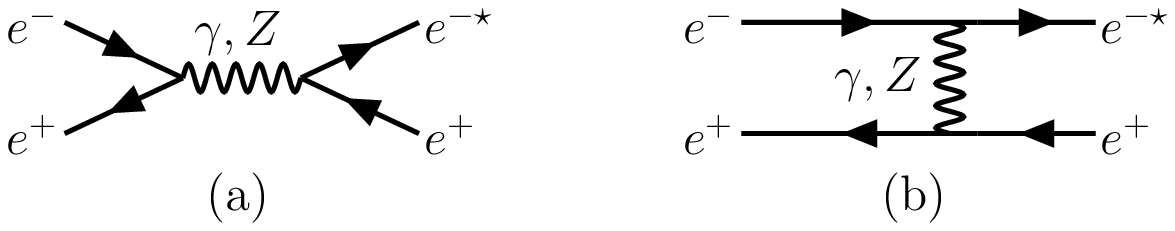}

\caption{Excited electron production at $e^{-}e^{+}$ colliders through the
(a) $s-$channel and (b) $t-$ channel exchange diagrams.\label{fig3}}
\end{figure}

High energy electron-positron collisions constitute an excellent environment
for the search for excited leptons. We examine the single production
of excited electrons ($e^{\star}$) at future $e^{-}e^{+}$ colliders
with $\sqrt{s}=500$ GeV, through the process $e^{-}e^{+}\rightarrow e^{\pm\star}e^{\mp}\rightarrow e^{\pm}e^{\mp}\gamma$.
The Feynman diagram for the process $e^{-}e^{+}\rightarrow e^{-\star}e^{+}$
is shown in Fig. \ref{fig3}.

\begin{figure}
\includegraphics{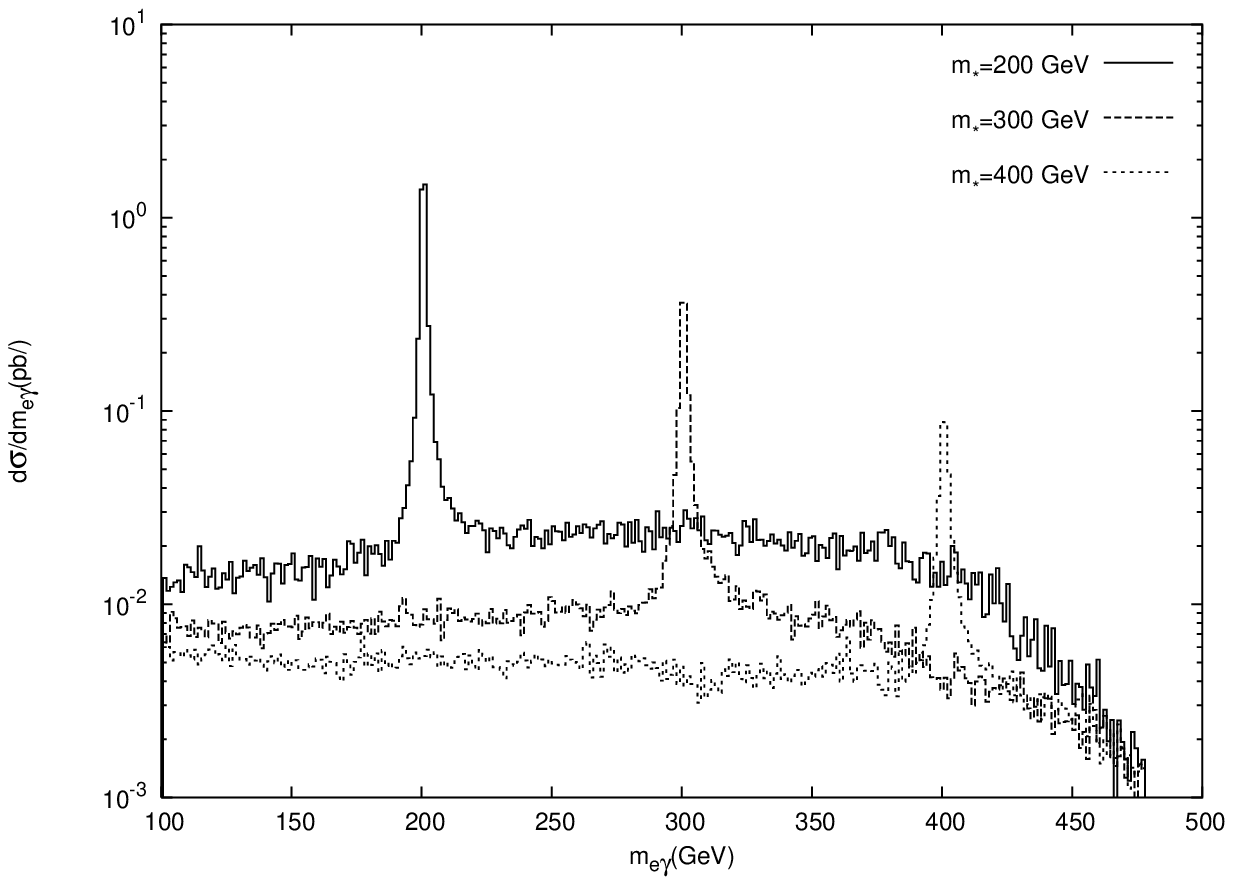}

\caption{Invariant mass $m_{e\gamma}$ distribution of signal (the scale $\Lambda=m_{*}$
and the coupling parameters $f=f^{'}=1$) and background at $e^{-}e^{+}$
collider.\label{fig4}}
\end{figure}

We applied to this process the following acceptance cuts

\begin{equation}
p_{T}^{e,\gamma}>20\textrm{ GeV}\end{equation}

\begin{equation}
|\eta_{e^{\pm},\gamma}|<2.5\end{equation}

\begin{equation}
\Delta R_{(e^{+}e^{-}),(e^{\pm}\gamma)}>0.4
\end{equation} where
$p_{T}$ is the transverse momentum of the visible particle. $\eta$
stands for the pseudo-rapidity of the visible particles and
$\Delta R=\sqrt{\Delta\eta^{2}+\Delta\phi^{2}}$ is the separation
between two of them. After applying these cuts, the SM background
cross section is found to be $\sigma=1.93$ pb. The $e\gamma$ decay
products of excited electron can be easily identified since they
typically have large transverse momentum of about $m_*/2$. Fig.
\ref{fig4} shows the invariant mass $m_{e\gamma}$ distribution in
the reaction $e^{+}e^{-}\rightarrow e^{+}e^{-}\gamma$ for the SM
and with the inclusion of an excited electron with masses
$m_{\star}=200$ GeV, $m_{\star}=300$ GeV, $m_{\star}=400$ GeV and
parameter$f=f^{'}=1$.

\begin{figure}
\includegraphics{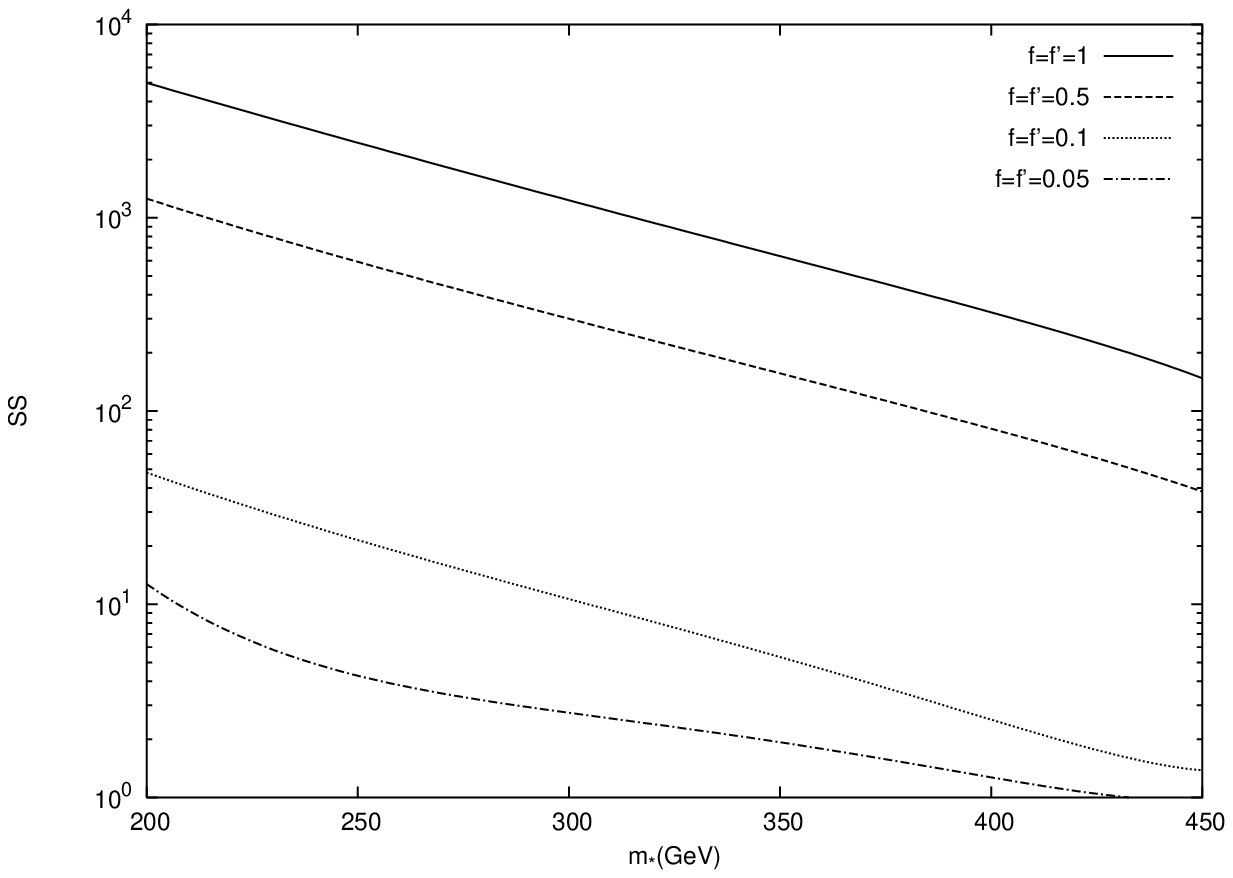}

\caption{Statististical significance depending on the excited electron mass
and different coupling parameters at the LC for the scale $\Lambda=m_{*}$.\label{fig5}}
\end{figure}

A natural way to extract the excited electron signal, and at the same
time suppress the SM backgrounds, is to impose a cut on the $e\gamma$
invariant mass. Therefore, we introduced the cut

\begin{equation}
|m_{e^{\pm}\gamma}-m_{\star}|<25\textrm{ GeV}\end{equation} for
considered excited electron mass range. In Table \ref{table1}, we
have presented the signal (for $f=f^{^{'}}=1$) and background
cross sections in $e\gamma$ invariant mass bins since the signal
is concentrated in a small region proportional to the invariant
mass resolution. In order to examine the potential of the collider
to search for the excited electron, we defined the statistical
significance $SS$ of the signal

\begin{equation}
SS=\frac{\left|\sigma_{S+B}-\sigma_{B}\right|}{\sqrt{\sigma_{B}}}\sqrt{L_{int}}
\end{equation}
where $L_{int}$ is the integrated luminosity of the collider. The
values of $SS$ evaluated at each excited electron mass points are
shown in the last column of Table \ref{table1}. As seen from the
Table \ref{table1} the calculated $SS$ values are higher than 5 up
to the center of mass energy of the LC. Single production of
excited electrons is dominated by the t-channel $\gamma$ exchange
contribution which makes its detection feasible up to masses next
to the $e^+e^-$ collider center of mass energy even with fairly
small magnetic transition couplings to electrons. For various
coupling parameters $f(=f^{'})$, we give the $SS$ values in Fig.
\ref{fig5}. Concerning the criteria above (SS>5), even for smaller
coupling as $f=f^\prime=0.1$ excited electrons with masses up to
375 GeV can be probed at the LC.

\begin{table}

\caption{The cross sections of the excited electron signal and relevant backgrounds
after the cuts at $e^{-}e^{+}$collider with $\sqrt{s}=0.5$ TeV and
$L_{int}=10^{5}$fb$^{-1}$ assuming $\Lambda=m_{\star}$ and $f=f'=1$.\label{table1}}

\begin{tabular}{|c|c|c|c|}
\hline
$m^{\star}$(GeV)&
$\sigma_{S+B}$(pb)&
$\sigma_{B}$(pb)&
$SS$\tabularnewline
\hline
200&
$5.93\times10^{0}$&
$1.35\times10^{-1}$&
4992.0\tabularnewline
\hline
250&
$3.27\times10^{0}$&
$1.68\times10^{-1}$&
2395.8\tabularnewline
\hline
300&
$1.95\times10^{0}$&
$2.02\times10^{-1}$&
1226.9\tabularnewline
\hline
350&
$1.14\times10^{0}$&
$2.26\times10^{-1}$&
606.9\tabularnewline
\hline
400&
$6.49\times10^{-1}$&
$1.86\times10^{-1}$&
185.4\tabularnewline
\hline
475&
$3.17\times10^{-1}$&
$1.16\times10^{-1}$&
81.3\tabularnewline
\hline
\end{tabular}
\end{table}

\subsection{$ep$ Collider}
\begin{figure}
\includegraphics{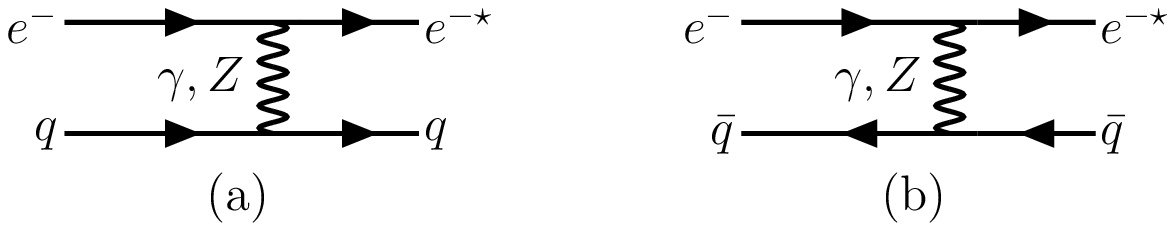}

\caption{Single production of excited electrons at $ep$ colliders through
the subprocesses (a) $e^{-}q\rightarrow e^{-\star}q$ and (b) $e^{-}\overline{q}\rightarrow e^{-\star}\overline{q}$.\label{fig6}}
\end{figure}

\begin{figure}
\includegraphics{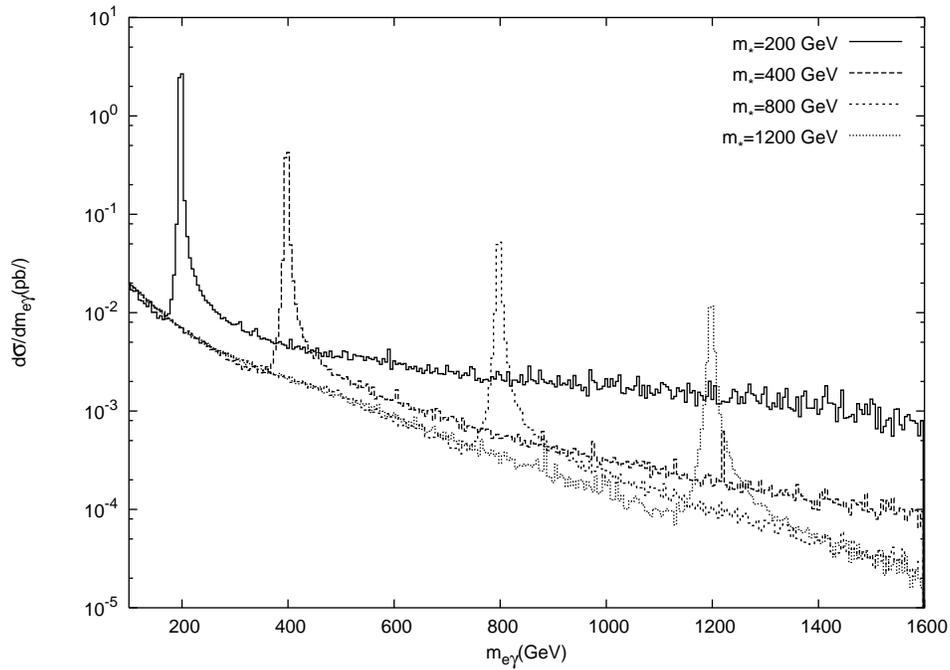}

\caption{Invariant mass $m_{e\gamma}$ distribution of signal ($\Lambda=m_{\star}$
and $f=f^{'}=1$) and background at $e^{-}p$ colliders.\label{fig7}}
\end{figure}

The magnetic transition couplings of excited electron to the
electron allows single production of $e^\star$ through $t$-channel
$\gamma$ and $Z$ exchange. The Feynman diagrams for the subprocess
$e^{-}q\rightarrow e^{-\star}q$ and $e^{-}\overline{q}\rightarrow
e^{-\star}\overline{q}$ are shown in Fig. \ref{fig6}.
After the acceptance cuts the total SM background cross section is
obtained as $\sigma_{B}=4.29$ pb. Fig. \ref{fig7} shows the invariant
mass $m_{e\gamma}$ distribution in the reaction $e^{-}q\rightarrow e^{-}\gamma q$
for the SM background and the signal (for $f=f^{^{'}}=1$) with the
inclusion of an excited electron with masses $m_{\star}=200$ GeV,
$m_{\star}=400$ GeV, $m_{\star}=800$ GeV and $m_{\star}=1200$ GeV.

\begin{figure}
\includegraphics{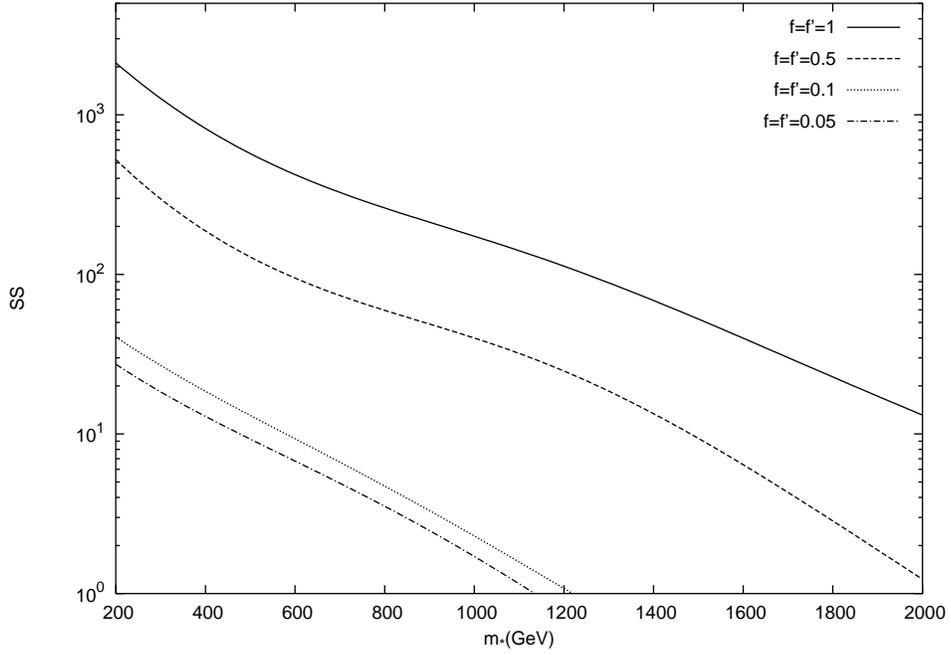}

\caption{Statististical significance depending on the excited electron mass
for different coupling parameters for the process$e^{-}p\rightarrow e^{-\star}q(\bar{q})X$
at the LC$\otimes$LHC.\label{fig8}}
\end{figure}

In Table \ref{table2}, we present the signal and background cross
sections in $e\gamma$ invariant mass bins satisfying the condition
$|m_{e^{-}\gamma}-m_{\star}|<25\textrm{ GeV}$ for the mass range
$m_{\star}=200-1200$ GeV and
$|m_{e^{-}\gamma}-m_{\star}|<50\textrm{ GeV}$ for
$m_{\star}=1200-2500$ GeV. For various coupling parameters
$f(=f^{'})$, we show the mass dependence of the $SS$ in Fig.
\ref{fig8}. As can be seen from Table \ref{table2} LC$\otimes$LHC
can discover excited electron in $e^{\star}\rightarrow e\gamma$
decay mode for $f=f^{'}=1$ up to the mass of 2300 GeV. This limit
is larger than the corresponding one which can be reached at 500
GeV $e^+e^-$ collider.

A common feature of the $e^+e^-$ and $ep$ collisions with respect
to the single production of excited electron is the prominent role
played by $t$-channel photon exchange mechanism which generates
large production rates. This also leads to the strong limits on
the compositeness scale in case of the negative search.

\begin{table}

\caption{The cross sections of the signal and relevant backgrounds at $e^{-}p$
collider with $\sqrt{s}=3.74$ TeV and $L_{int}=100$ pb$^{-1}$ assuming
$\Lambda=m_{\star}$ and $f=f'=1$. \label{table2}}

\begin{tabular}{|c|c|c|c|}
\hline
Mass, GeV&
$\Delta\sigma_{S+B}$(pb)&
$\Delta\sigma_{B}$(pb)&
SS\tabularnewline
\hline
200&
$1.55\times10^{1}$&
$5.03\times10^{-1}$&
2117.5\tabularnewline
\hline
400&
$1.32\times10^{0}$&
$1.02\times10^{-1}$&
696.5\tabularnewline
\hline
600&
$6.68\times10^{-1}$&
$3.37\times10^{-2}$&
345.4\tabularnewline
\hline
800&
$2.78\times10^{-1}$&
$1.25\times10^{-2}$&
237.1\tabularnewline
\hline
1000&
$1.25\times10^{-1}$&
$4.44\times10^{-3}$&
181.6\tabularnewline
\hline
1200&
$6.23\times10^{-2}$&
$1.83\times10^{-3}$&
141.3\tabularnewline
\hline
1500&
$1.82\times10^{-2}$&
$1.17\times10^{-3}$&
50.2\tabularnewline
\hline
2000&
$5.63\times10^{-3}$&
$1.16\times10^{-3}$&
13.1\tabularnewline
\hline
2500&
$2.01\times10^{-3}$&
$1.16\times10^{-3}$&
2.5\tabularnewline
\hline
\end{tabular}
\end{table}

\subsection{$pp$ Collider}

\begin{figure}
\includegraphics{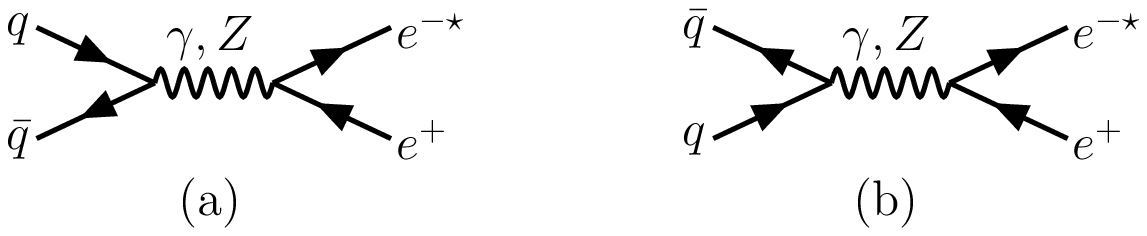}

\includegraphics{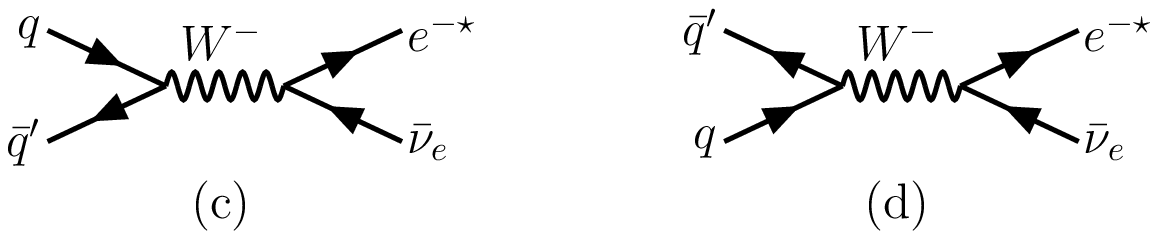}

\caption{Excited electron production via the (a,b) photon and $Z-$boson exchange,
and (c,d) $W^{-}-$boson in the $s-$channel diagrams at hadron colliders.\label{fig9}}
\end{figure}

At the LHC, the single production of excited electrons takes place
through the subprocesses $q\bar{q}\rightarrow Z/\gamma\rightarrow e^{\pm}e^{\mp\star}\rightarrow e^{\pm}e^{\mp}\gamma$
and $q\bar{q}^{'}\rightarrow W^{\mp}\rightarrow\nu e^{\mp\star}\rightarrow\nu e^{\mp}\gamma$
via the Drell-Yan mechanism. The diagrams related to these subprocesses
are shown in Fig. \ref{fig9}. After the acceptance cuts the total
SM background cross sections are obtained as $\sigma_{B}=1.16$ pb
for the process $pp\rightarrow e^{+}e^{-}\gamma X$ and $\sigma_{B}=1.35$
pb for $pp\rightarrow\nu e^{-}\gamma X$.

\begin{figure}
\includegraphics{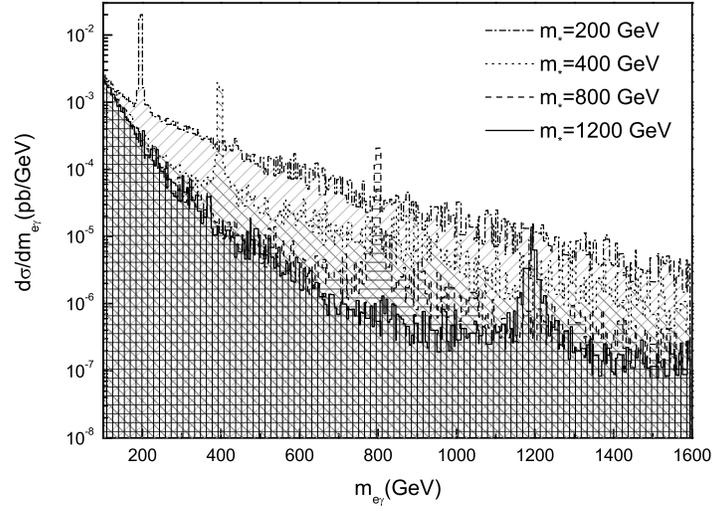}

\begin{center}(a)\end{center}

\includegraphics{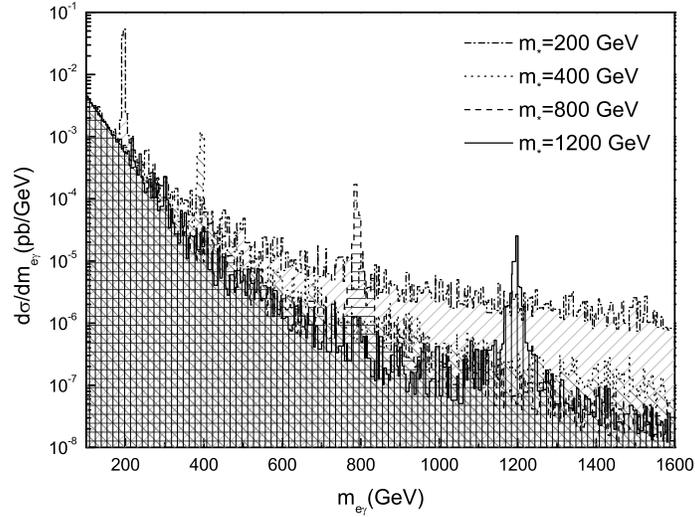}

\begin{center}(b)\end{center}

\caption{Invariant mass $m_{e\gamma}$ distribution of signal and background
for the processes (a) $pp\rightarrow e^{+}e^{-}\gamma X$ and (b)
$pp\rightarrow\nu e^{-}\gamma X$ at $pp$ collider (LHC).\label{fig10}}
\end{figure}

\begin{figure}
\includegraphics{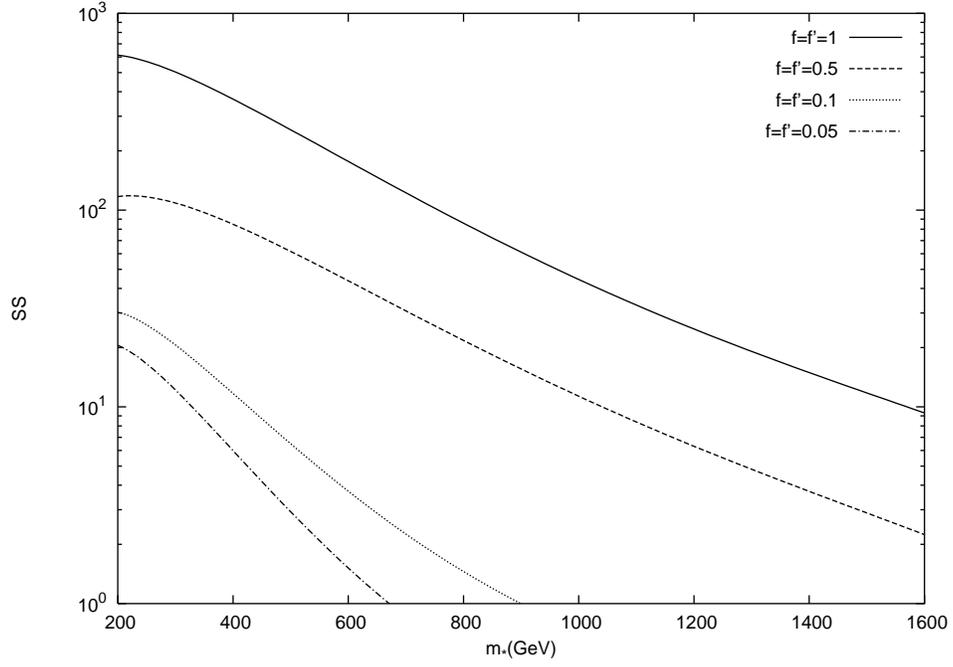}

\begin{center}(a)\end{center}

\includegraphics{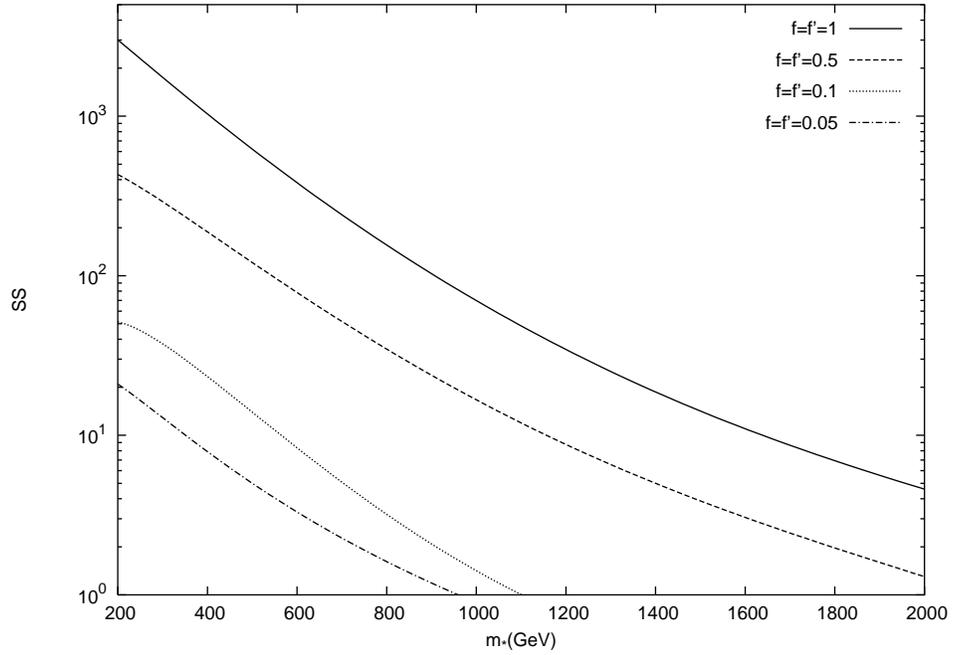}

\begin{center}(b)\end{center}

\caption{Statistical significance depending on the excited electron mass for
process (a) $pp\rightarrow e^{+}e^{-}\gamma X$ and (b) $pp\rightarrow\nu e^{-}\gamma X$
with different couplings $f(=f')$ at the LHC.\label{fig11}}
\end{figure}

Fig. \ref{fig10} shows the invariant mass $m_{e\gamma}$
distributions in the reactions $pp\rightarrow e^{+}e^{-}\gamma X$
and $pp\rightarrow\nu e^{-}\gamma X$ for the SM background and the
signal (for $f=f^{^{'}}=1$) with the inclusion of an excited
electron with masses $m_{\star}=200$ GeV, $m_{\star}=400$ GeV,
$m_{\star}=800$ GeV and $m_{\star}=1200$ GeV. In Table
\ref{table3}, we present the signal and background cross sections
in $e\gamma$ invariant mass bins satisfying the condition
$|m_{e^{-}\gamma}-m_{\star}|<25\textrm{ GeV}$ for excited electron
mass $m_{\star}=200-1200$ GeV and
$|m_{e^{-}\gamma}-m_{\star}|<50\textrm{ GeV}$ for
$m_{\star}=1200-2500$ GeV. Statistical significance SS are shown
in Fig. \ref{fig11} (a) for $pp\rightarrow e^{+}e^{-}\gamma X$ and
(b) for $pp\rightarrow\nu e^{-}\gamma X$ processes with different
couplings $f(=f')$. One can conclude that the results obtained in
this study turns out to be at least an order of magnitude more
stringent than the present best limits coming from the HERA
experiments. Moreover, the LHC will be able to extend considerably
the range of excited electron masses that can be probed (up to
about 2 TeV).

\begin{table}

\caption{The cross sections of the signal and relevant backgrounds at $pp$
collider (LHC) with $\sqrt{s}=14$ TeV and $L_{int}=10^{5}$pb$^{-1}$.
The statistical significance $SS$ are given for the coupling $f=f'=1$
and the scale $\Lambda=m^{\star}$. \label{table3} }

\begin{tabular}{|c|c|c|c|c|c|c|}
\hline
Final states$\rightarrow$&
\multicolumn{3}{c|}{$e^{-}e^{+}\gamma$}&
\multicolumn{3}{c|}{$e^{-}\overline{\nu}\gamma$}\tabularnewline
\hline
Mass, GeV &
$\sigma_{S+B}$(pb)&
$\sigma_{B}$(pb)&
SS&
$\sigma_{S+B}$(pb)&
$\sigma_{B}$(pb)&
SS\tabularnewline
\hline
200&
$2.33\times10^{-1}$&
$1.29\times10^{-2}$&
613.6&
$2.98\times10^{-1}$&
$9.79\times10^{-4}$&
3003.8\tabularnewline
\hline
400&
$2.24\times10^{-2}$&
$1.79\times10^{-4}$&
524.4&
$2.52\times10^{-2}$&
$5.98\times10^{-5}$&
1027.5\tabularnewline
\hline
800&
$1.67\times10^{-3}$&
$8.01\times10^{-5}$&
56.1&
$1.65\times10^{-3}$&
$2.28\times10^{-5}$&
107.7\tabularnewline
\hline
1200&
$3.13\times10^{-4}$&
$1.56\times10^{-5}$&
23.8&
$2.91\times10^{-4}$&
$1.29\times10^{-5}$&
24.4\tabularnewline
\hline
1600&
$7.63\times10^{-5}$&
$5.75\times10^{-6}$&
9.3&
$7.00\times10^{-5}$&
$5.81\times10^{-6}$&
8.4\tabularnewline
\hline
2000&
$2.26\times10^{-5}$&
$1.24\times10^{-6}$&
6.1&
$1.97\times10^{-5}$&
$1.46\times10^{-6}$&
4.8\tabularnewline
\hline
2500&
$6.27\times10^{-6}$&
$7.35\times10^{-7}$&
2.0&
$4.97\times10^{-6}$&
$6.14\times10^{-7}$&
1.7\tabularnewline
\hline
\end{tabular}
\end{table}

\section{Conclusion}

We give the realistic estimates for excited electron signal and
the corresponding background at three-type of colliders with the
availability of higher center of mass energies and higher
luminosities. Since the cross section for the signal is
proportional to $1/\Lambda^{2}$, various choices of the $\Lambda$,
i.e. in this study we have choosen $\Lambda=m_{\star}$, will lead
to the changes in the cross sections as $(m_{\star}/\Lambda)^{2}.$
For $\Lambda=1$ TeV, we need to multiply the signal cross sections
by a factor {[}$m_{\star}$(TeV){]}$^{2}$ at every mass values of
excited electrons. This factor also extends the attainable mass
limits for $m_{\star}>1$ TeV. Our analysis show that for
$f=f^{'}=1$ LC can discover excited electron in
$e^{\star}\rightarrow e\gamma$ decay mode up to the kinematical
limit, while LC$\otimes$LHC and LHC can reach much higher mass
values, namely 2300 GeV and 1900 GeV, respectively. For
$f=f^{^{'}}=0.05$ discover limits are: 240 GeV at LC, 650 GeV at
LC$\otimes$LHC and 450 GeV at LHC.

In our analysis, we assumed that the excited electron interact
with the SM particles via the effective Lagrangian (1). In
principle, excited electron may also couple to ordinary quarks and
leptons via contact interactions which can enlarge discovery
limits for $pp$ colliders \cite{9,13} as well as for $ep$
colliders. However, we restrict ourselves to the gauge
interactions since the aim of this paper is to compare the
potential of three types of colliders within the similar sets of
cuts. Our results coincide with \cite{10} where similar analysis
was performed for the LC, and essentially coincide with \cite{14}
where different parton distribution functions, namely MRS (G)
\cite{15}, have been used for the single production of excited
electrons at the LHC. Finally, our analysis show that
LC$\otimes$LHC is more promising than the LHC and much more
promising than the LC for the processes considered.

\begin{acknowledgments}
This work is partially supported by Turkish State Planning Committee
under the Grants No 2002K120250 and 2003K120190.
\end{acknowledgments}


\begin{thebibliography}{99}
\bibitem{1}S. Sultansoy, Eur. Phys. J. C (2004) DOI 10.1140/epjcd/s2004-03-1716-2;
hep-ex/0306034.
\bibitem{2}H. Terazawa, Y. Chikashige and K. Akama, Phys. Rev. D \textbf{15},
480 (1977); Y. Ne'eman, Phys. lett. B \textbf{82}, 69 (1979); H. Terazawa,
M. Yasue, K. Akama and M. Hayashi, Phys. Lett. B \textbf{112}, 387
(1982).
\bibitem{3}F. M. Renard, Il Nuovo Cimento A \textbf{77}, 1 (1983);
E.J. Eichten, K.D. Lane and M.E. Peskin, Phys. Rev. D \textbf{50},
811 (1983); A. De Rujula, L. Maiani and R. Petronzio, Phys. Lett. B
\textbf{140}, 253 (1984); J. Kühn and P.M. Zerwas, Phys. Lett. B
147, 189 (1984).
\bibitem{4}K. Hagiwara et al., Particle Data Group, Phys. Rev. D \textbf{66}, 010001 (2002).
\bibitem{5}M. Acciarri et al., L3 Coll., Phys. Lett. B \textbf{502}, 37 (2001).
\bibitem{6}C. Adloff et al., H1 Coll., Eur. Phys. J. C \textbf{17}, 567 (2000).
\bibitem{7}P. Achard et al., L3 Coll., Phys. Lett. B \textbf{531}, 28 (2002).
\bibitem{8}K. Hagiwara, S. Komamiya, and D. Zeppenfeld, Z. Phys. C \textbf{29}, 115 (1985).
\bibitem{9}U. Baur, M. Spira and P.M. Zerwas, Phys. Rev. D \textbf{42}, 815 (1990).
\bibitem{10}F. Boudjema, A. Djouadi and J.L. Kneur, Z. Phys. C \textbf{57}, 425
(1993); F. Boudjema, A. Djouadi, Phys. Lett. B \textbf{240}, 485
(1990).
\bibitem{11}A. Pukhov \emph{et al.}, hep-ph/9908288 (1999).
\bibitem{12}H.L. Lai et al., CTEQ Collab., Phys. Rev. D \textbf{51}, 4763 (2000).
\bibitem{13}O. Cakir \emph{et al.}, Eur. Phys. J. C \textbf{30}, d01, 005 (2003).
\bibitem{14}O.J.P. Eboli, S.M. Lietti and P. Mathews, Phys. Rev. D \textbf{65}, 075003 (2002).
\bibitem{15}A. D. Martin, W.J. Stirling and R.G. Roberts, Phys. Lett. B \textbf{354}, 155
(1995).
\end{thebibliography}
\end{document}